\documentclass[conference]{IEEEtran}
\IEEEoverridecommandlockouts
\usepackage{cite}
\usepackage{amsmath,amssymb,amsfonts}
\usepackage{algorithmic}
\usepackage{booktabs} 
\usepackage{multirow} 
\usepackage{multicol}
\usepackage{graphicx}
\usepackage{textcomp}
\usepackage{xcolor}
\def\BibTeX{{\rm B\kern-.05em{\sc i\kern-.025em b}\kern-.08em
    T\kern-.1667em\lower.7ex\hbox{E}\kern-.125emX}}

\newcommand{\linebreakand}{%
  \end{@IEEEauthorhalign}
  \hfill\mbox{}\par
  \mbox{}\hfill
  \begin{@IEEEauthorhalign}
}

\begin{document}

\title{DS-ViT: Dual-Stream Vision Transformer for Cross-Task Distillation in Alzheimer’s Early Diagnosis\\

}

\author{\IEEEauthorblockN{Ke Chen}
\IEEEauthorblockA{\textit{School of Information Sciences} \\
\textit{University of Illinois Urbana-Champaign}\\
Urbana, USA \\
kec10@illinois.edu}
\and
\IEEEauthorblockN{Yifeng Wang \qquad}
\IEEEauthorblockA{\textit{Department of Electrical and
Computer Engineering \qquad}\\
\textit{Carnegie Mellon University \qquad}\\
Pittsburgh, USA \qquad \\
yifengw3@andrew.cmu.edu \qquad}
\and
\IEEEauthorblockN{ \qquad \qquad \qquad \qquad Yufei Zhou}
\IEEEauthorblockA{\textit{ \qquad \qquad \qquad \qquad Department of Economics} \\
\textit{ \qquad \qquad \qquad \qquad Duke University}\\
 \qquad \qquad \qquad \qquad Durham, USA \\
 \qquad \qquad \qquad \qquad yz597@duke.edu.com}
\and
\IEEEauthorblockN{ \qquad \qquad Haohan Wang}
\IEEEauthorblockA{\textit{ \qquad \qquad School of Information Sciences} \\
\textit{ \qquad \qquad University of Illinois Urbana-Champaign}\\
 \qquad \qquad Urbana, USA \\
 \qquad \qquad haohanw@illinois.edu}
}

\maketitle

\begin{abstract}
In the field of Alzheimer’s disease diagnosis, segmentation and classification tasks are inherently interconnected. Sharing knowledge between models for these tasks can significantly improve training efficiency, particularly when training data is scarce. However, traditional knowledge distillation techniques often struggle to bridge the gap between segmentation and classification due to the distinct nature of tasks and different model architectures. To address this challenge, we propose a dual-stream pipeline that facilitates cross-task and cross-architecture knowledge sharing. Our approach introduces a dual-stream embedding module that unifies feature representations from segmentation and classification models, enabling dimensional integration of these features to guide the classification model. We validated our method on multiple 3D datasets for Alzheimer’s disease diagnosis, demonstrating significant improvements in classification performance, especially on small datasets. Furthermore, we extended our pipeline with a residual temporal attention mechanism for early diagnosis, utilizing images taken before the atrophy of patients' brain mass. This advancement shows promise in enabling diagnosis approximately six months earlier in mild and asymptomatic stages, offering critical time for intervention.
\end{abstract}


\section{Introduction}

Alzheimer’s Disease (AD) remains one of the most challenging neurodegenerative disorders, primarily affecting the elderly. Despite extensive research, the current therapeutic approaches are largely symptomatic, aiming to manage cognitive and behavioral symptoms rather than offering a cure \cite{b1}. The primary drugs available, including donepezil, memantine, galantamine, and rivastigmine, target neurotransmitter systems to temporarily stabilize cognitive functions \cite{b2}. However, their effectiveness is limited during the later stages of the disease when significant neurodegeneration has already occurred. This has led to a growing recognition of the need for early diagnosis and intervention in the early asymptomatic stage, which could potentially slow or halt the progression of AD by targeting the disease in its preclinical stages \cite{b3}.

\begin{figure}
\centering
\includegraphics[scale=0.25]{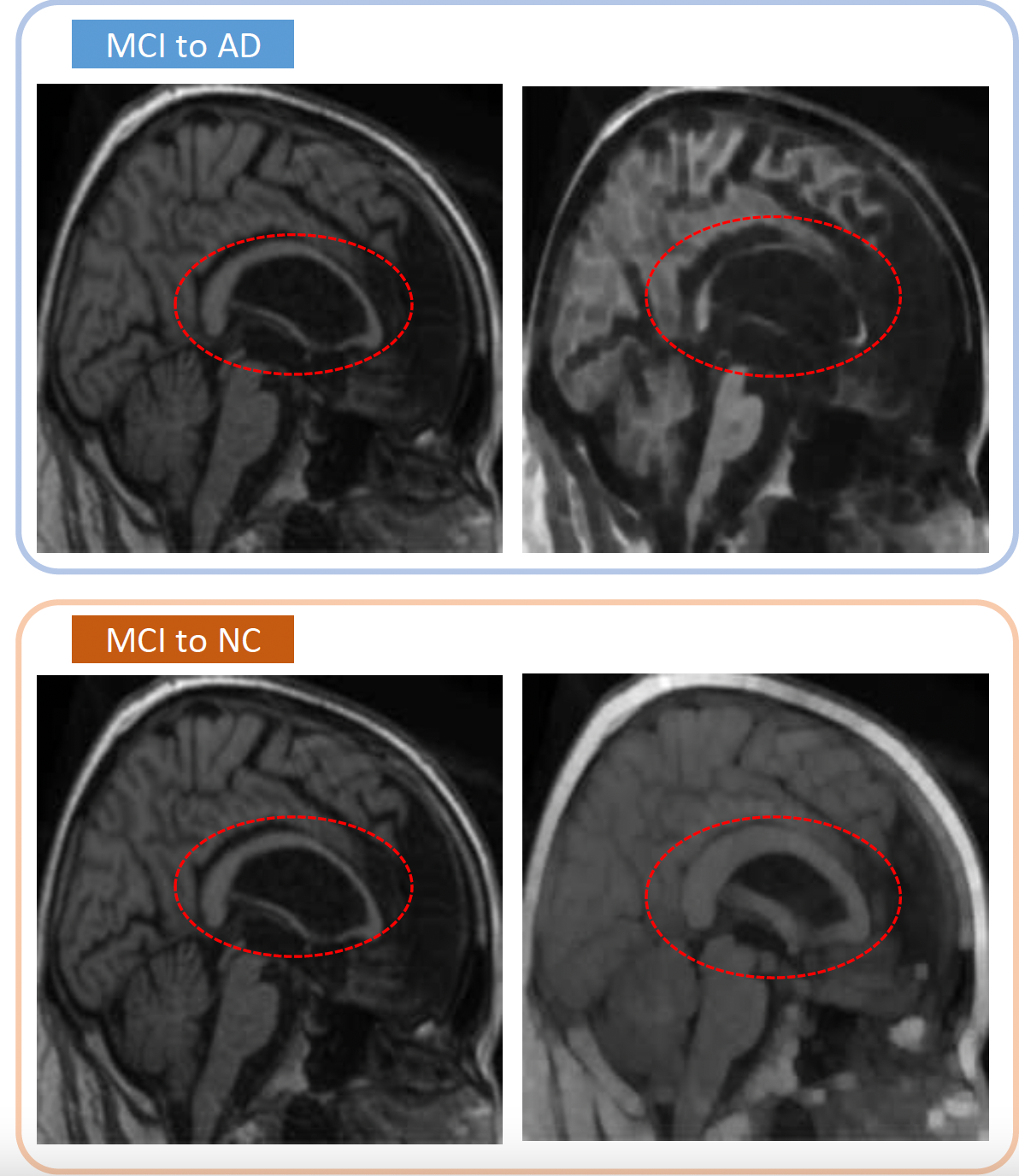}
  \caption{MRI scans illustrating the volumetric changes in the cerebral ventricle (highlighted in red) during the transition from Mild Cognitive Impairment (MCI) to Alzheimer’s Disease (AD) and Normal Case (NC). The progression from MCI to AD is characterized by a noticeable increase in ventricle size, while the reverse trend is observed in MCI to NC \cite{b3}. These structural alterations provide evidence of detectable early-stage biomarkers, supporting the feasibility of AI-driven early diagnosis of Alzheimer’s disease. 
  }
  \label{trend}
\end{figure}

Magnetic Resonance Imaging (MRI) is a crucial tool in the diagnosis of AD, providing high-resolution images that reveal structural changes in the brain, such as the atrophy of brain regions and the enlargement of the cerebral ventricles \cite{b4}. These changes are closely associated with the progression of AD and can serve as important biomarkers for early detection \cite{b3}. Fig. \ref{trend} illustrates the volumetric changes in the cerebral ventricle during the transition from Mild Cognitive Impairment (MCI) to AD, highlighting the potential of these biomarkers for AI-driven early diagnosis.

Traditional approaches in machine-learning-based AD diagnosis have often treated segmentation and classification as separate tasks. This separation overlooks the critical medical connection between the anatomical segmentation of brain structures and the subsequent classification of disease progression. Accurate segmentation provides essential insights into the structural changes associated with AD, such as atrophy in specific brain regions, which are directly linked to the severity and advancement of the disease \cite{b5}. However, when segmentation and classification are handled independently, it can lead to inefficiencies in training and suboptimal performance, particularly when high-quality AD imaging data is scarce and expensive to obtain.

These limitations significantly hinder the application of ML in AD diagnosis. Without integrated segmentation and classification, ML models struggle to capture the full complexity of AD’s progression, resulting in less accurate predictions and a reduced ability to identify early-stage biomarkers. This is particularly problematic in clinical settings, where the ability to detect AD early is crucial for timely intervention and treatment. The lack of effective integration between segmentation and classification tasks in current ML approaches directly impacts the reliability and clinical applicability of these models.

FastSurfer and ADAPT represent state-of-the-art models in brain MRI segmentation and classification, respectively, offering vast opportunities for improving AD diagnosis. FastSurfer, a well-established CNN-based MRI segmentation model trained on large-scale datasets, excels in detailed brain segmentation, which is crucial for identifying the structural changes associated with AD \cite{b6}. On the other hand, ADAPT, a Vision Transformer (ViT)-based model, has demonstrated remarkable performance in AD diagnosis with minimal parameters \cite{b7}.

Given their complementary strengths, there is a significant potential for knowledge sharing between these models to enhance diagnostic accuracy. Knowledge distillation, a technique that transfers knowledge from a robust teacher model to a more efficient student model, could provide a direct pathway for this information exchange. However, traditional knowledge distillation methods, which rely heavily on soft labels \cite{b8}, are inadequate for tasks with fundamentally different objectives, such as segmentation and classification. Moreover, the architectural differences between FastSurfer and ADAPT pose additional challenges to effective knowledge transfer.

To overcome these challenges, we propose the Dual-Stream Vision Transformer (DS-ViT) pipeline, a novel approach that integrates segmentation and classification tasks to enhance training efficiency and model accuracy in the diagnosis of AD. Our approach utilizes FastSurfer as the teacher model to provide detailed brain segmentation, and ADAPT as the student model, leveraging its strong performance in AD diagnosis. To bridge the gap between these two tasks, we designed a Dual-Stream Embedding module that processes both pixel-level MRI image data and token-like segmentation data, embedding and integrating them through a 3D Bottleneck MLP structure. This innovative design facilitates effective cross-task knowledge sharing and improves the model’s ability to generalize from limited data.

As shown in Fig. \ref{trend}, the progression from Mild Cognitive Impairment (MCI) to Alzheimer's Disease (AD) is characterized by subtle yet critical changes in brain structure, such as the volumetric increase in the cerebral ventricle. These patterns, observable through sequential MRI scans, suggest that AI has the potential to detect early biomarkers of AD, thereby enabling timely intervention before significant symptoms emerge. The ability to predict disease progression at an early stage could significantly improve patient outcomes by allowing for preventive treatments.

To harness this potential and extend the application of our methods, we incorporated a Residual Temporal Attention Block (RTAB) into the DS-ViT pipeline. The RTAB is designed to capture temporal dynamics by analyzing differences between feature maps from sequential MRI scans. By aggregating these temporal features through an attention mechanism, the RTAB enables the model to predict the risk of disease progression, thereby supporting early intervention strategies aimed at delaying or preventing the onset of AD.

We rigorously evaluated the DS-ViT pipeline across multiple benchmarks, demonstrating its superiority over the baseline ADAPT model. Our results indicate a significant improvement in classification accuracy, particularly in scenarios with limited training data, where DS-ViT achieved an average accuracy increase of 7\% and reduced convergence time by half. Furthermore, the RTAB-enhanced DS-ViT pipeline achieved a 70\% overall prediction accuracy, with an accuracy of 86\% for high-confidence samples, indicating not only enhanced diagnostic accuracy but also the potential for asymptomatic stage intervention in high-risk patients.

The contributions of this work are threefold: (1) We introduce the DS-ViT pipeline, which successfully integrates segmentation knowledge into a classification framework, (2) We demonstrate the effectiveness of the Dual-Stream Embedding and 3D Bottleneck MLP structure in enhancing model performance, and (3) We extend the model’s application to early diagnosis, offering a novel approach to predicting Alzheimer’s disease progression with clinically meaningful lead time.

\section{Related Works}
\subsection{Single Modality AD Diagnosis}
Single-modality approaches, particularly those utilizing Magnetic Resonance Imaging (MRI), have been extensively studied for Alzheimer's Disease (AD) diagnosis. These methods focus on analyzing volumetric brain data to detect structural changes such as cortical atrophy and ventricular enlargement, which are hallmark features of AD \cite{b5}. Various deep learning models, including Uni4Eye, I3D, MedicalNet, 3D ResNet, and 3D DenseNet, have been developed to leverage MRI data for this purpose \cite{b10,b11,b12,b13,b14}. These models predominantly employ convolutional neural networks (CNNs) to capture spatial and temporal patterns in the brain, aiming to improve diagnostic accuracy and robustness.

Despite the successes of these single-modality models, they often face limitations in fully capturing the complex and heterogeneous nature of AD, particularly when data is scarce or when subtle structural changes need to be detected early. The reliance solely on volumetric data may overlook critical contextual information that could enhance diagnosis. This has led to the exploration of multi-modality approaches, which integrate additional data sources—such as segmentation results or other imaging modalities—to improve performance.

The DS-ViT pipeline proposed in this work builds on these single-modality approaches by integrating segmentation knowledge from FastSurfer with traditional MRI-based classification. This integration aims to overcome some of the limitations of single-modality models, providing a more comprehensive and accurate diagnostic tool for AD, particularly in early-stage detection.

\subsection{Knowledge Distillation}
In the field of visual AI, knowledge distillation methods are commonly employed to share knowledge between models, thereby enhancing the performance of the student model. Traditional knowledge distillation primarily relies on soft labels generated by the teacher model, which replace the conventional hard labels \cite{b8}. These soft labels reflect the probability distribution of each class as predicted by the teacher model, offering a more nuanced view than hard labels, which assign a binary classification. The student model learns from these soft labels to approximate the predictive capability of the teacher model. However, the fundamental differences between segmentation and classification tasks limit the applicability of traditional knowledge distillation methods. In segmentation tasks, the model’s output are pixel-level labels, whereas classification tasks focus on the label for the entire image. The differences in output format and task objectives make it difficult to apply soft label-based knowledge distillation directly to these tasks. Moreover, the challenge is exacerbated by the architectural differences between the teacher and student models. Common knowledge distillation variants, such as hint layer distillation \cite{b15} and structural model distillation \cite{b16}, usually assume a certain level of similarity in feature representation between the teacher and student models. In our task, this challenge is intensified by the distinct architectures of the teacher and student models (CNN and Transformer), making cross-task knowledge distillation even more complex.

In this work, we address these limitations of traditional knowledge distillation by proposing the DS-ViT pipeline. Our approach achieves cross-task and cross-architecture knowledge sharing, breaking down the barriers between different tasks and model architectures.

\subsection{FastSurfer}
In scenarios where training data is limited, leveraging knowledge from models trained on large-scale datasets can significantly enhance training efficiency and performance. FastSurfer, an extensively validated deep-learning pipeline, is designed for the fully automated processing of structural human brain MRIs \cite{b6}. This pipeline is known for its robust performance and speed, processing each MRI scan in less than one minute. FastSurfer excels in two main areas: detailed brain segmentation and surface reconstruction.

The brain segmentation component of FastSurfer is particularly noteworthy due to its precision. It divides the brain into 95 distinct anatomical regions, a level of detail that surpasses many other segmentation tools. This granularity is crucial for research and clinical applications, as understanding the structure and function of different brain areas is essential. FastSurfer’s segmentation results have been extensively validated against established neuroimaging protocols, demonstrating high accuracy and reliability across diverse datasets.

We believe that the detailed segmentation provided by FastSurfer is exceptionally well-suited to guide Alzheimer’s disease early diagnosis. Understanding the intricate differences between various brain regions is a critical prerequisite for accurate diagnosis. By incorporating FastSurfer’s segmentation outputs into our diagnostic pipeline, we can enrich the student model’s training with refined anatomical knowledge, thereby improving its predictive performance on smaller, disease-specific datasets.

\subsection{ADAPT}
In our previous work, we developed Alzheimer’s Diagnosis through Adaptive Profiling Transformers (ADAPT), which efficiently detects Alzheimer's disease by converting  3D brain MRIs into 2D slices along different planes (axial, coronal, and sagittal) \cite{b7}. Using dimension-specific self-attention and cross-attention mechanisms, ADAPT captures critical spatial relationships within the data. Despite its small parameter size, ADAPT demonstrated exceptional performance in Alzheimer’s diagnosis, outperforming several state-of-the-art 3D image classification models, including MedicalNet and 3D DenseNet across various datasets.

In this work, we build upon ADAPT by integrating FastSurfer’s detailed brain segmentation knowledge, enhancing performance, especially with limited training data. This integration allows ADAPT to utilize anatomical insights crucial for early Alzheimer’s diagnosis, improving accuracy and extending the model’s applicability to more complex diagnostic tasks. Notably, the proposed enhancements are adaptable to other Transformer-based feature extractors, such as 3D Vision Transformer. This adaptability offers broader applicability in medical image analysis.

\begin{figure*}[htbp]
    \centering
    \includegraphics[width=0.8\textwidth]{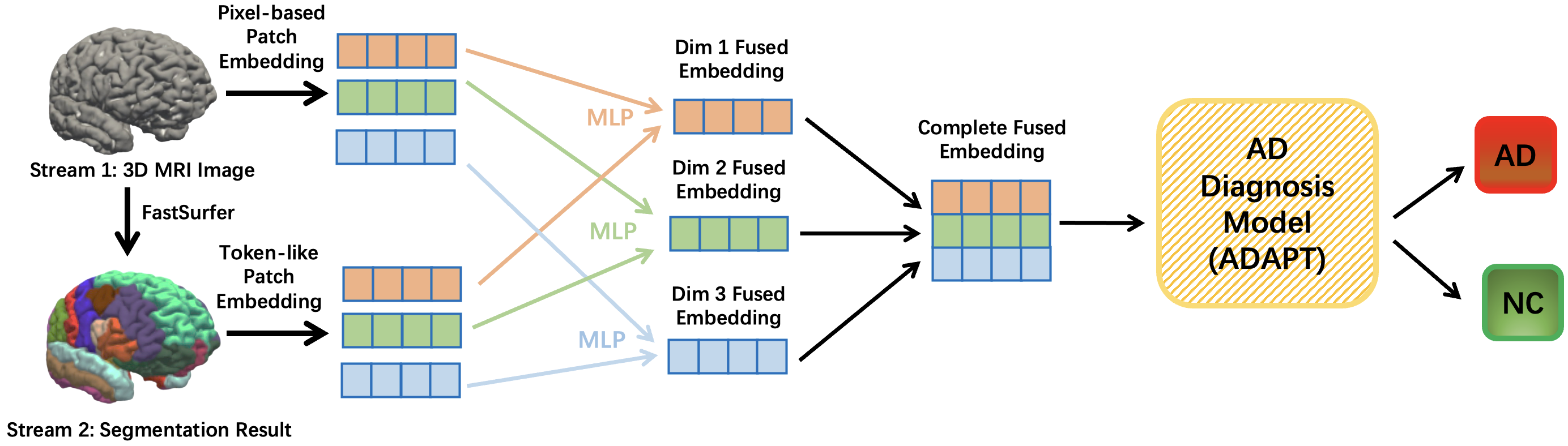}
    \caption{The process starts with generating segmentation results from 3D MRI images using FastSurfer. These images and their segmentation maps are sliced along three orthogonal planes. Stream 1 processes pixel-based image data, while Stream 2 handles token-like embeddings from the segmentation results. The embedded features from both streams are integrated using separate trainable MLP modules across three dimensions, then concatenated into a comprehensive feature matrix. The matrix serves as the input to the AD diagnosis model, i.e. ADAPT in our setup, which performs feature extraction and diagnosis.}
    \label{pipeline}    
\end{figure*}

\section{Methods}
DS-ViT is a 3D image classification pipeline built upon ADAPT, a baseline model known for its strong performance in small parameter size. Unlike conventional biomedical imaging models that rely solely on pathological images as input, DS-ViT is designed to process dual-stream inputs. One stream captures pixel-level information from the original MRI images, while the other incorporates segmentation results from FastSurfer. FastSurfer, a widely validated 3D MRI segmentation model trained on large-scale datasets, encodes extensive knowledge about brain regions that are crucial for the accurate diagnosis of Alzheimer’s disease. By leveraging FastSurfer as a teacher model, we aim to enhance the training efficiency and diagnostic accuracy of ADAPT in Alzheimer’s detection.

Our approach consists of two main components. The first is the design of the DS-ViT structure, tailored for diagnosing Alzheimer’s disease from single 3D MRI images. The second component extends the functionality of DS-ViT by utilizing multiple time-point scans from the same patient to predict future disease risk. This extension enables the early detection and proactive treatment of Alzheimer’s disease, offering the possibility of earlier intervention for patients.
\begin{figure*}[htbp]
    \centering
    \includegraphics[width=\textwidth]{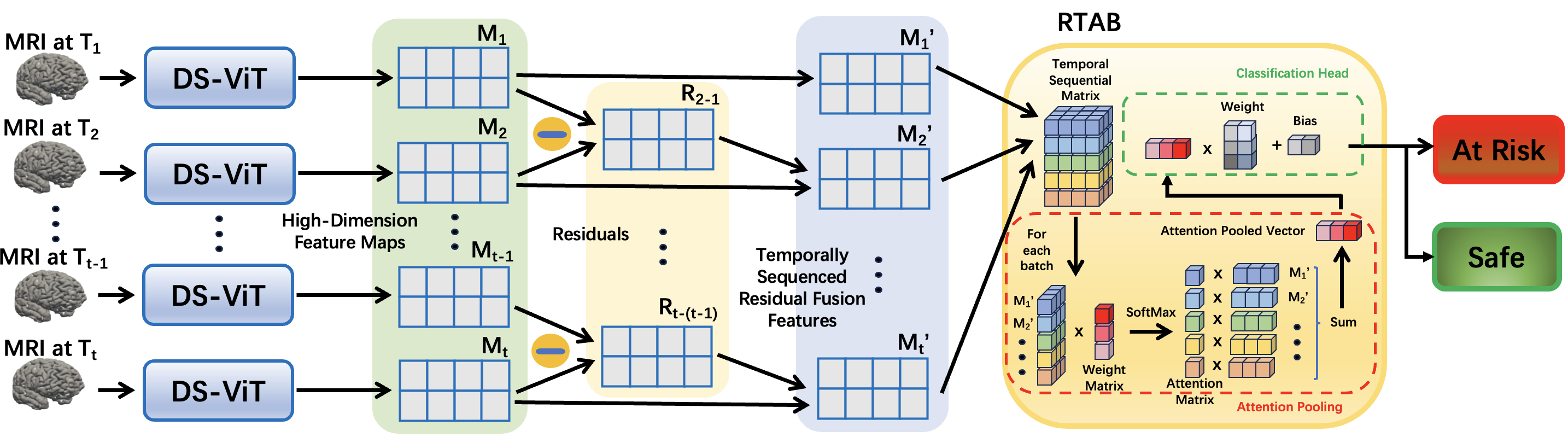}
    \caption{The overview of the pipeline for DS-ViT + RTAB for Alzheimer's Early Diagnosis. DS-ViT extracts high-dimensional feature maps from MRI scans at different time points. Residuals are computed between consecutive feature maps and fused with the current map to create residual fusion features. These features are aggregated by the Residual Temporal Attention Block (RTAB) to assess the risk of Alzheimer’s disease progression, outputting “At Risk” or “Safe” classifications for early intervention.}
    \label{ED_pipeline}    
\end{figure*}

\subsection{DS-ViT Pipeline}
As shown in Fig. \ref{pipeline}, we first process the 3D MRI images using FastSurfer to obtain segmentation results that contain detailed brain region information. The original MRI images and the corresponding segmentation maps are then sliced along three orthogonal planes, and embedded with the dual-stream embedding module, which will be detailed in  section \ref{embedding}. The resulting embedding vectors from both streams, across the three dimensions, are then integrated through three trainable MLP modules. These integrated features are concatenated to form the input for ADAPT, which subsequently learns from this fused matrix and make predictions.
\subsubsection{Dual-Stream Embedding}
\label{embedding}
Since ADAPT operates by slicing 3D MRI images along three orthogonal planes and then embedding and concatenating the slices, our Dual-Stream Embedding is essentially embedding these 2D slices.

\textbf{Stream 1: Pixel-based Patch Embedding for MRI Images}\\
This step is based on the traditional ViT approach to image embedding. The image is first divided into fixed-size patches. Each patch undergoes convolutional operations to extract features and resize them to a fixed dimension, which are then merged. To preserve spatial information, positional embeddings are added to each patch’s feature matrix, ensuring that the positional context is maintained during the embedding process.

\textbf{Stream 2: Token-like Patch Embedding for Segmentation Results}\\
Unlike MRI images, where pixel values are continuous, segmentation results contain discrete class labels representing different brain regions. We have tested and proved that applying convolutional operations directly to these segmentation maps can lead to significant issues, where pixel-based patch embeddings proved ineffective. Recognizing the categorical nature of segmentation data, we treat each brain region label as a unique token, similar to tokens in Natural Language Processing (NLP).

For each patch $\mathbf{S}_p$ in the segmentation map, let each pixel $s_i \in \mathbf{S}_p$ take a discrete value from a set of 95 classes, $s_i \in \{1, 2, \dots, 95\}$. Each class $s_i$ is mapped to a learnable embedding vector $\mathbf{e}_i \in \mathbb{R}^{D}$ using an embedding matrix $\mathbf{E} \in \mathbb{R}^{K \times D}$:

\[
\mathbf{z}_i = \mathbf{E}[s_i]
\]

where $\mathbf{z}_i \in \mathbb{R}^{D}$ is the embedding vector for the pixel $s_i$. 

To obtain the feature representation for the entire patch $\mathbf{S}_p$, we first aggregate the embeddings of all pixels within the patch through feature concatenation and then apply average pooling:

\[
\mathbf{z}_p = \frac{1}{|\mathbf{S}_p|} \sum_{i=1}^{|\mathbf{S}_p|} \mathbf{z}_i
\]

where $|\mathbf{S}_p|$ is the number of pixels in the patch. The resulting $\mathbf{z}_p \in \mathbb{R}^{D}$ represents the token-like embedding vector for the patch.

To maintain the effectiveness and consistency of positional information, the same patch division mechanism is used as in the pixel-based embedding, and positional embeddings are shared with those used for the MRI images. This ensures that spatial context is preserved across both streams.

\subsubsection{3D Feature Integration}
To effectively fuse the features from the two streams, we employ a trainable Bottleneck MLP structure. This structure consists of multiple fully connected layers, which first reduce the dimensionality of the features and then restore them to a specified size. Given that ADAPT processes slices from three orthogonal dimensions independently, and considering the significant differences between slices from different orientations, we initialize a separate MLP module for each dimension. This approach ensures the independent training of MLPs across the three dimensions, preserving the distinct characteristics of each slice orientation.

After the features are integrated within each dimension, they are concatenated along the three dimensions to form a complete feature matrix. This fused matrix serves as the input to the ADAPT encoder, where ADAPT completes the subsequent high-dimensional feature extraction and classification tasks.

\subsection{Extension: Early Diagnosis of Alzheimer's Disease}

The basic DS-ViT Pipeline is designed to diagnose whether a patient is already afflicted with Alzheimer's disease, meaning that diagnosis only occurs once the disease is present, offering no possibility for early intervention. To extend the application of our work and provide greater benefits to patients, we have enhanced DS-ViT with a Residual Temporal Attention Mechanism. This addition enables our method to predict future disease risk based on a series of early MRI images from a patient, thus allowing for the possibility of preventive measures and earlier intervention.

\subsection{DS-ViT for Early Diagnosis Pipeline}

In the Early Diagnosis process, we retain most of the DS-ViT structure but replace the classification head with a Residual Temporal Attention Block (RTAB), as shown in Fig. \ref{ED_pipeline}. In this setup, DS-ViT functions as a feature extractor, processing multiple MRI images taken at different time points to generate a sequence of high-dimensional feature maps. Each feature map $M_t$ is subtracted from the feature map $M_{t-1}$ from the previous time point to obtain the residual $R_t$. This residual is then fused with $M_t$ through an MLP to produce the residual fusion feature $M_t'$. For the earliest feature map $M_1$, since there is no prior feature map to compute the residual, we define its residual fusion feature as itself, i.e., $M_1' = M_1$. These temporally sequenced residual fusion features are then input into the RTAB, which synthesizes the early features and trends to predict the risk of deterioration at a future examination time point.

\begin{table*}[htbp]
\renewcommand{\arraystretch}{1.1}
\centering
\caption{Comparison of Models Across Data-Sufficient and Data-Scarce Scenarios}
\label{baseline}
\begin{tabular}{@{}lccccccc@{}}
\toprule
\multirow{2}{*}{\textbf{Model Name}} & \multicolumn{4}{c}{\textbf{Data-Sufficient Scenario}} & \multicolumn{3}{c}{\textbf{Data-Scarce Scenario}} \\ \cmidrule(lr){2-5} \cmidrule(l){6-8}
 & \textbf{ADNI (val)} & \textbf{MIRIAD} & \textbf{AIBL} & \textbf{OASIS} & \textbf{MIRIAD (val)} & \textbf{AIBL} & \textbf{OASIS} \\ \midrule
\textbf{DS-ViT} & \textbf{0.933} & \textbf{0.941} & \textbf{0.914} & \textbf{0.859} & \textbf{0.899} & \textbf{0.892} & \textbf{0.823} \\ 
\textbf{ADAPT} & 0.924 & 0.903 & 0.910 & 0.817 & 0.829 & 0.815 & 0.761 \\ 
\textbf{Uni4Eye} & 0.646 & 0.647 & 0.697 & 0.688 & 0.617 & 0.626 & 0.583 \\ 
\textbf{I3D} & 0.629 & 0.448 & 0.674 & 0.607 & 0.549 & 0.513 & 0.472 \\ 
\textbf{MedicalNet-10} & 0.843 & 0.847 & 0.856 & 0.793 & 0.802 & 0.782 & 0.694 \\ 
\textbf{MedicalNet-34} & 0.669 & 0.782 & 0.847 & 0.585 & 0.592 & 0.643 & 0.551 \\ 
\textbf{MedicalNet-101} & 0.425 & 0.656 & 0.320 & 0.291 & 0.547 & 0.592 & 0.415 \\ 
\textbf{3D ResNet-34} & 0.618 & 0.453 & 0.779 & 0.745 & 0.581 & 0.647 & 0.422 \\ 
\textbf{3D ResNet-50} & 0.618 & 0.657 & 0.776 & 0.720 & 0.543 & 0.483 & 0.621 \\ 
\textbf{3D ResNet-101} & 0.526 & 0.426 & 0.601 & 0.433 & 0.505 & 0.574 & 0.450 \\ 
\textbf{3D DenseNet-121} & 0.675 & 0.408 & 0.816 & 0.772 & 0.724 & 0.398 & 0.536 \\ 
\textbf{3D DenseNet-201} & 0.612 & 0.553 & 0.672 & 0.671 & 0.578 & 0.630 & 0.427 \\ 

\bottomrule
\end{tabular}
\end{table*}

\subsection{Residual Temporal Attention Block (RTAB)}

To effectively aggregate these temporal features, we propose an attention-based feature aggregation module. This module utilizes an AttentionPooling mechanism to perform weighted aggregation of the residual fusion features across different time points.

Given an input sequence of residual fusion features $X = [\mathbf{M}_1', \mathbf{M}_2', \dots, \mathbf{M}_{T}']$ with $T$ time steps, each feature $\mathbf{M}_t' \in \mathbb{R}^{d}$ where $d$ is the dimensionality of the features, the input sequence is first reshaped to separate the batch size and sequence length. This reshaped sequence is represented as:

\[
X \in \mathbb{R}^{\text{batch\_size} \times T \times d}
\]

Next, the attention weights for each time step are computed using a linear transformation:

\[
\mathbf{w}_t = \text{softmax}\left(\mathbf{W}_a \mathbf{M}_t' \right)
\]

where $\mathbf{W}_a \in \mathbb{R}^{d \times 1}$ is a learnable weight matrix, and $\mathbf{w}_t \in \mathbb{R}$ represents the attention weight for the $t$-th time step. The softmax function ensures that the attention weights sum to 1 across all time steps.

These attention weights are then used to compute a weighted sum of the input features:

\[
\mathbf{P} = \sum_{t=1}^{T} \mathbf{w}_t \mathbf{M}_t'
\]

where $\mathbf{P} \in \mathbb{R}^{d}$ is the pooled feature vector, effectively capturing the temporal dynamics of the input sequence. 

This pooled feature vector $\mathbf{P}$ is then passed through a classification head, implemented as a fully connected layer, to produce the final prediction logits:

\[
\text{logits} = \mathbf{W}_c \mathbf{P} + \mathbf{b}_c
\]

where $\mathbf{W}_c \in \mathbb{R}^{\text{num\_classes} \times d}$ is the weight matrix and $\mathbf{b}_c \in \mathbb{R}^{\text{num\_classes}}$ is the bias term.

The RTAB module leverages the temporal variation in patient features, helping the model to more accurately predict the risk of future deterioration. The attention-based feature aggregation plays a crucial role in our method, significantly enhancing classification performance, especially in scenarios with limited data.

\section{Experiments}

\subsection{Dataset and Preprocessing}
To validate the effectiveness of the DS-ViT Pipeline, we designed two experimental scenarios. The first scenario simulates a situation with ample training data, using the Alzheimer’s Disease Neuroimaging Initiative (ADNI) dataset for training and the Minimal Interval Resonance Imaging in Alzheimer's Disease (MIRIAD) , Open Access Series of Imaging Studies (OASIS), and Australian Imaging, Biomarker and Lifestyle Flagship Study of Ageing (AIBL) \cite{b17,b18,b19,b20} datasets for testing. The second scenario simulates a data-scarce environment, using the MIRIAD dataset for training and the OASIS and AIBL datasets for testing.

We framed the task as a binary classification task. In certain datasets where three categories are present—Alzheimer’s Disease (AD), Mild Cognitive Impairment (MCI), and Normal Controls (NC)—we excluded the MCI samples, focusing only on AD and NC cases. Evatually, the large training dataset, ADNI, comprises 1,216 NC and 1,110 AD samples. In the small training dataset, MIRIAD, after balancing, we retained 354 NC and 346 AD samples. Since the MIRIAD dataset includes multiple scans from the same patients with identical diagnostic outcomes, we partitioned the training and validation sets at the patient level to mitigate the risk of overfitting and data leakage due to individual patient variability. For the test datasets, AIBL contains 363 NC and 50 AD samples, while OASIS includes 1,692 NC and 465 AD samples.

In the context of the early diagnosis task, due to the constraints of computational resources, we initially set the temporal sequence length to 2, meaning each subject contributes two prior MRI scans. However, this sequence length can be extended to improve accuracy if computational resources allow. All patients for this task were selected from the ADNI dataset. If a patient, initially diagnosed as non-AD, exhibited disease progression in subsequent scans (e.g., NC$\rightarrow$MCI, NC$\rightarrow$AD, MCI$\rightarrow$AD), they were classified as at-risk. Conversely, if there was no disease progression (e.g., consistently NC, consistently MCI, or MCI$\rightarrow$NC), they were classified as safe. After filtering and balancing, we obtained 279 AD and 276 NC patients, with each patient contributing two sequential MRI scans.

All MRI images mentioned above underwent consistent data preprocessing techniques to ensure data quality and format uniformity. Specifically, each MRI image was first subjected to bias field correction using the N4ITK algorithm to address intensity inconsistencies \cite{b21}. The images were then aligned to the MNI space via affine registration, using the SyN algorithm from ANTs with the ICBM 2009c nonlinear symmetric template \cite{b22}. Background regions were removed from the registered images, resulting in a consistent voxel size of 1 mm isotropic. A deep quality control system was employed to verify the registration accuracy, and images with an accuracy below 0.5 were excluded from further analysis \cite{b23}.

\begin{table*}[htbp]
\renewcommand{\arraystretch}{1.1}
\centering
\caption{Ablation Study Results in Data-Scarce Scenario}
\label{ablation}
\begin{tabular}{@{}lcccccc@{}}
\toprule
\multirow{2}{*}{\textbf{Model Name}} & \multicolumn{2}{c}{\textbf{MIRIAD (val)}} & \multicolumn{2}{c}{\textbf{AIBL}} & \multicolumn{2}{c}{\textbf{OASIS}} \\ \cmidrule(lr){2-3} \cmidrule(lr){4-5} \cmidrule(l){6-7}
 & \textbf{Acc} & \textbf{Recall} & \textbf{Acc} & \textbf{Recall} & \textbf{Acc} & \textbf{Recall} \\ \midrule
\textbf{DS-ViT} & \textbf{0.899} & \textbf{0.917} & \textbf{0.892} & \textbf{0.879} & \textbf{0.823} & \textbf{0.847} \\ 
\textbf{Hint Layer Distillation} & 0.583 & 0.539 & 0.443 & 0.466 & 0.315 & 0.404 \\ 
\textbf{ADAPT (W/O Seg Stream)} & 0.829 & 0.856 & 0.815 & 0.824 & 0.761 & 0.802 \\ 
\textbf{W/O MRI Stream} & 0.794 & 0.824 & 0.733 & 0.790 & 0.702 & 0.684 \\ 
\textbf{W/O Dual-stream Emb} & 0.647 & 0.661 & 0.638 & 0.644 & 0.572 & 0.508 \\ 
\bottomrule
\end{tabular}
\end{table*}

\subsection{Evaluation}
\subsubsection{Evaluation on DS-ViT Pipeline}

We conducted a series of experiments to assess the performance improvements of the DS-ViT pipeline over baseline models and to validate the effectiveness of each component within our approach. Two sets of experiments were designed to evaluate the model under data-sufficient and data-scarce scenarios.

In the first set of comparative experiments, we evaluated the performance of DS-ViT against several baseline models, including ADAPT, Uni4Eye, I3D, MedicalNet, 3D ResNet, and 3D DenseNet \cite{b10,b11,b12,b13,b14}. These models were tested under two conditions: a data-sufficient scenario where the models were trained on the ADNI dataset, and a data-scarce scenario where the models were trained on the MIRIAD dataset. The results of these experiments are presented in Table \ref{baseline}.

In the second set of ablation studies, five experiments were conducted to determine the necessity of each component in our proposed approach:

\begin{itemize}
\item \textbf{DS-ViT}: The control group, used to evaluate the performance improvements achieved by DS-ViT.
\item \textbf{Hint Layer Distillation}: Applied a traditional distillation method to assess the necessity of our innovations.
\item \textbf{ADAPT}: Simulated a single-stream input scenario without the segmentation input, to evaluate the impact of excluding segmentation data.
\item \textbf{W/O MRI Stream}: Simulated a single-stream input scenario without the MRI image input, to test the importance of including MRI data.
\item \textbf{W/O Dual-Stream Embedding}: Applied pixel-based embedding to both streams before feature fusion, to validate the advantages of the dual-stream embedding strategy.
\end{itemize}

Accuracy and recall were used as the primary evaluation metrics, reflecting the overall classification performance and the model’s sensitivity to detecting diseased samples, respectively. The results of these experiments are detailed in Table \ref{ablation}.

From the data presented in Table \ref{baseline}, it is evident that DS-ViT consistently outperforms the baseline models across all datasets, demonstrating the effectiveness of our approach. Notably, when trained on small datasets (e.g., MIRIAD), DS-ViT achieved a performance improvement of 7\% over its base model, ADAPT. Additionally, DS-ViT typically converges within approximately 15 epochs, whereas ADAPT requires 30 to 60 epochs to reach convergence. These results suggest that the DS-ViT pipeline significantly enhances both the accuracy and training efficiency of the baseline model, with the improvements being particularly pronounced under conditions of limited training data.

Furthermore, as shown in the ablation study results in Table \ref{ablation}, DS-ViT achieved a 33\% higher accuracy compared to the traditional distillation method (Hint Layer Distillation), highlighting the effectiveness of our innovation. The comparison between DS-ViT and the single-stream input groups (W/O MRI Stream and W/O Segmentation Stream) indicates that the dual-stream fusion consistently yields better results than using either stream alone, thereby validating the necessity of cross-task learning through dual-stream input. Similarly, the comparison between DS-ViT and the W/O Dual-Stream Embedding experiment confirms that our dual-stream embedding strategy, specifically designed for handling different types of inputs, is highly effective.

\subsubsection{Evaluation on Early Diagnosis}
For the early diagnosis task, since no existing models have been designed to incorporate temporal sequences of multiple MRI images for this purpose, we primarily rely on our experimental results to assess the performance. We conducted four experiments:
\begin{itemize}
	\item \textbf{DS-ViT+RTAB (Experimental Group)}: This configuration combines DS-ViT with RTAB to evaluate the performance of our enhanced pipeline.
	\item \textbf{DS-ViT (Single-Timepoint)}: This variant uses only the most recent MRI scan to predict the diagnosis of the next examination, allowing us to assess the necessity of incorporating multiple time-point inputs.
	\item \textbf{DS-ViT+SVM}: DS-ViT is used as the feature extractor, with the extracted high-dimensional features classified using a Support Vector Machine (SVM), to validate the superiority of RTAB over traditional classifiers.
	\item \textbf{DS-ViT+ResNet}: Similarly, DS-ViT is used as the feature extractor, but the extracted features are classified using a ResNet model, further testing the effectiveness of RTAB.
\end{itemize}

\begin{table}[htbp]
\centering
\caption{Ablation Experiments to Validate the Superiority of RTAB}
\label{RATB}
\begin{tabular}{|l|c|}
\hline
\textbf{Model Name} & \textbf{Accuracy} \\ \hline
\textbf{DS-ViT+RTAB} & \textbf{0.704} \\ \hline
\textbf{DS-ViT} & 0.560 \\ \hline
\textbf{DS-ViT+SVM} & 0.603 \\ \hline
\textbf{DS-ViT+ResNet} & 0.614 \\ \hline
\end{tabular}
\end{table}
As shown in Table \ref{RATB}, the DS-ViT+RTAB configuration achieved the highest accuracy among all tested methods and ablation experiments, demonstrating the superiority of the DS-ViT pipeline when combined with the RTAB module, as well as the potential for early prediction of Alzheimer’s disease. However, the overall accuracy was only 70.4\%, prompting a deeper analysis of the DS-ViT+RTAB results. We found that the majority of misclassifications occurred in samples where the model’s confidence was relatively low. When we focused on high-confidence samples—those with predicted probabilities of 70–100\% for one class and 0–30\% for the other—the accuracy increased to 86\%. This suggests that our model can provide reliable recommendations for patients in cases where it exhibits high confidence.

Upon further analysis, we found that for the selected 555 patients, the average interval between the two MRI scans used was approximately 6 months. This indicates that when our model confidently identifies a patient as at risk, it offers the possibility of initiating treatment up to 6 months earlier, allowing for preventive measures and interventions before the disease significantly progresses.

\section{Conclusions}
In this paper, we proposed the Dual-Stream Vision Transformer (DS-ViT) pipeline, a novel approach designed to integrate segmentation and classification tasks for enhanced performance in Alzheimer’s disease diagnosis. By leveraging the robust segmentation capabilities of FastSurfer and incorporating a dual-stream embedding mechanism, our approach successfully bridges the gap between distinct tasks and architectures, resulting in improved classification accuracy and training efficiency, particularly in data-scarce scenarios.

Our experiments demonstrated that DS-ViT consistently outperforms the baseline models across various datasets, achieving significant gains in accuracy while reducing convergence time. The introduction of the Residual Temporal Attention Block (RTAB) further extends the application of DS-ViT to early diagnosis, offering a clinically valuable tool for predicting Alzheimer’s disease progression up to 6 months in advance. Our methods not only improve diagnostic accuracy but also provide a critical window for early intervention, potentially altering the course of the disease for high-risk patients.

The contributions of this work are multifaceted: (1) the successful integration of segmentation knowledge into a classification framework, (2) the validation of dual-stream embedding and 3D Bottleneck MLP structures as effective components for cross-task knowledge sharing, and (3) the extension of these techniques to the early diagnosis of Alzheimer’s disease, providing the opportunities for prevention and early treatment.

Future work will focus on further enhancing the temporal modeling capabilities of DS-ViT by increasing the sequence length of input MRI scans and exploring the application of our pipeline to other neurodegenerative diseases. Additionally, we aim to refine our attention mechanisms to better handle low-confidence cases, thus broadening the practical applicability of our model in diverse clinical settings.

\bibliography{conference}

\vspace{12pt}

\end{document}